\documentclass[11pt,a4paper]{article}
\pdfoutput=1
\usepackage{jheppub}
\usepackage{soul}
\usepackage{graphicx}
\usepackage{amsmath}
\usepackage{amsfonts}
\usepackage{mathrsfs}
\usepackage{amssymb}
\usepackage[utf8]{inputenc}
\usepackage{multicol}
\usepackage{array}
\usepackage{slashed}
\usepackage{afterpage}

\newcommand{\ds}[1]{{\displaystyle #1 }}
\def\lsim{\raise0.3ex\hbox{$\;<$\kern-0.75em\raise-1.1ex
\hbox{$\sim\;$}}}
\def\gsim{\raise0.3ex\hbox{$\;>$\kern-0.75em\raise-1.1ex
\hbox{$\sim\;$}}}

\title{Dark Matter and Exotic Neutrino Interactions\\in Direct Detection Searches}
\author[a]{Enrico Bertuzzo,}
\author[b]{Frank F. Deppisch,}
\author[c]{Suchita Kulkarni,}
\author[a]{Yuber F. Perez Gonzalez,}
\author[a]{Renata Zukanovich Funchal}

\emailAdd{bertuzzo@if.usp.br}
\emailAdd{f.deppisch@ucl.ac.uk}
\emailAdd{suchita.kulkarni@oeaw.ac.at}
\emailAdd{yfperezg@if.usp.br}
\emailAdd{zukanov@if.usp.br}

\affiliation[a]{Departamento de F\'isica Matem\'atica, Instituto de F\'isica, Universidade de S\~ao Paulo,\\ 
R. do Mat\~ao 1371, CEP.\ 05508-090, S\~ao Paulo, Brazil}
\affiliation[b]{Department of Physics and Astronomy, University College London, \\ London WC1E 6BT, United Kingdom}
\affiliation[c]{Institut f\"ur Hochenergiephysik,  
\"Osterreichische Akademie der Wissenschaften, \\ Nikolsdorfer Gasse 18, 1050 Wien, Austria}

\abstract{We investigate the effect of new physics interacting with both Dark Matter (DM) and neutrinos at DM direct detection experiments. Working within a simplified model formalism, we consider vector and scalar mediators to determine the scattering of DM as well as the modified scattering of solar neutrinos off nuclei. Using existing data from LUX as well as the expected sensitivity of LUX-ZEPLIN and DARWIN, we set limits on the couplings of the mediators to quarks, neutrinos and DM. Given the current limits, we also assess the true DM discovery potential of direct detection experiments under the presence of exotic neutrino interactions. In the case of a vector mediator, we show that the DM discovery reach of future experiments is affected for DM masses $m_\chi \lesssim 10$~GeV or DM scattering cross sections $\sigma_\chi \lesssim 10^{-47}$~cm$^2$. On the other hand, a scalar mediator will not affect the discovery reach appreciably.}

\keywords{}

\begin{document}
\begin{flushright}
HEPHY-PUB 983/17
\end{flushright}

\maketitle

\section{Introduction}
\label{sec:intro}
The Standard Model (SM) of particle physics, despite its enormous success in describing experimental data, cannot explain DM observations. This has motivated a plethora of Beyond the Standard Model (BSM) extensions. Despite intense searches, none of these BSM extensions have been experimentally observed, leaving us with little knowledge of the exact nature of DM.

The lack of an experimentally discovered theoretical framework that connects the SM degrees of freedom with the DM sector has led to a huge activity in BSM model building. Among various DM scenarios, the Weakly Interacting Massive Particle (WIMP) DM remains the most attractive one, with several experiments actively searching for signs of WIMPs. Within this paradigm, DM is a stable particle by virtue of a $\mathcal{Z}_2$ symmetry under which it is odd. The WIMP interactions with the SM particles can be detected via annihilation (at indirect detection experiments), production (at collider experiments) and scattering (at direct detection experiments). If the WIMP idea is correct, the Earth is subjected to a wind of DM particles that interact weakly with ordinary matter, thus direct detection experiments form a crucial component in the experimental strategies to detect them.

At direct detection experiments, WIMP interactions are expected to induce nuclear recoil events in the detector target material. These nuclear recoils can be, in most detectors, discriminated from the electron recoils produced by other incident particles. Depending on the target material and the nature of DM-SM interactions, two different kind of DM interactions can be probed: spin-dependent and spin-independent DM-nucleus scattering. The current limits for spin-independent DM-nucleus interactions are considerably more stringent, and the next generation of direct detection experiments will probe the spin-independent interactions even further by lowering the energy threshold and increasing the exposure.

A signal similar to DM scattering can also be produced by coherent neutrino scattering off nuclei (CNSN) in DM direct detection experiments~\cite{Cabrera:1984rr,Drukier:1986tm,Strigari:2009bq}, hence constituting a background to the WIMP signal at these experiments. Unlike more conventional backgrounds such as low energy electron recoil events or neutron scattering due to ambient radioactivity and cosmic ray exposure, the CNSN background can not be reduced. The main sources contributing to the neutrino background are the fluxes of solar and atmospheric neutrinos~\cite{Gutlein:2010tq}, both fairly well measured in neutrino oscillation experiments~\cite{Aharmim:2011vm, Wendell:2010md}.  Within the SM, CNSN originates from the exchange of a Z boson via neutral currents. Given the minuscule SM cross-section $\sigma_{\rm CNSN} \lsim 10^{-39}$ cm$^2$ for neutrino energies $\lsim 10$ MeV, and the insensitivity of the existing DM detectors to this cross-section, CNSN events have yet to be experimentally observed. The minimum DM -- nucleus scattering cross-section at which the neutrino background becomes unavoidable is termed the {\em neutrino floor}~\cite{Billard:2013qya}. In fact, the {\em neutrino floor} limits the DM discovery potential of direct detection experiments, so diminishing the uncertainties on the determination of solar and atmospheric neutrino fluxes as well as the direct measurement of CNSN is very important. Fortunately, dedicated experiments are being developed to try to directly detect CNSN~\cite{Moroni:2014wia,Akimov:2015nza} in the very near future.

The absence of any WIMP signal at the existing direct detection experiments has resulted in the need for next generation experiments. It is expected that these experiments will eventually reach the sensitivity to measure solar (and perhaps atmospheric) neutrinos from the {\em neutrino floor}. It thus becomes important to analyse the capacity of these experiments to discriminate between DM and neutrino scattering events. It has been shown that a sufficiently strong Non-Standard Interaction (NSI) contribution to the neutrino -- nucleus scattering can result in a signal at direct detection experiments \cite{Harnik:2012ni, Pospelov:2012gm, Cerdeno:2016sfi, Dent:2016wor}. Several attempts have been made to discriminate between DM scattering and neutrino scattering events~\cite{Dent:2016iht, Cerdeno:2016sfi, Franarin:2016ppr, Grothaus:2014hja, Dent:2016wor}.

The cases considered so far involve the presence of BSM in either the neutrino scattering or the DM sector. However, it is likely that a BSM mediator communicates with both the neutrino and the DM sector, and the DM and a hidden sector may even be responsible for the light neutrino mass generation \cite{Huang:2014bva}. In such cases, it is important to consider the combined effect of neutrino and DM scattering at direct detection experiments. In this work we analyse quantitatively the effect of the presence of BSM physics communicating to both the neutrinos and the DM sector on the DM discovery potential at future direct detection experiments.

The paper is organized as follows. In section~\ref{sec:frame} we define the simplified BSM models we consider while section~\ref{sec:dd} is dedicated to calculational details of neutrino scattering and DM scattering at direct detection experiments. Equipped with this machinery, in section~\ref{sec:limits}, we describe the statistical procedure used to derive constraints for the existing and future experiments. We consider the impact of the BSM physics in the discovery potential of direct detection experiments in section~\ref{sec:modified_spectra}. Finally, we conclude in section~\ref{sec:conc}.

\section{The framework}
\label{sec:frame}

Working within the framework of simplified models, we consider scenarios where the SM is extended with one DM and one mediator field. The DM particle is odd under a $\mathcal{Z}_2$ symmetry, while the mediator and the SM content is $\mathcal{Z}_2$ even. The symmetry forbids the decay of DM to SM particles and leads to $2 \rightarrow 2$ processes between SM and DM sector which results in the relic density generation, as well as signals at (in)direct detection experiments.

To be concrete, we extend the SM sector by a Dirac DM fermion, $\chi$, with mass $m_\chi$ and consider two distinct possibilities for the mediator. In our analysis we will only specify the couplings which are relevant for CNSN and DM-nucleus scattering, namely, the couplings of the mediator to quarks, neutrinos and DM. We will explicitly neglect mediator couplings to charged leptons, and comment briefly on possible UV-complete models.

\subsection{Vector mediator}
%
In this scenario, we extend the SM by adding a Dirac fermion DM, $\chi$, with mass $m_\chi$ and a real vector boson, $V_\mu$, with mass $m_V$. The relevant terms in the Lagrangian are:
\begin{align}
\label{eq:simp_model_vec}
{\cal L}_{\rm vec}  &= V_\mu (J_f^\mu + J_\chi^\mu)+ \displaystyle \frac{1}{2}\,m_V^2 V_\mu V^\mu \, , 
\end{align}
where the currents are ($f=\{u,d,\nu\}$)
\begin{align}
\label{eq:currents}
J_f^\mu = \sum_f  \overline{f} \gamma^\mu (g_V^f + g_A^f \gamma_5 ) f \, , \quad
J_\chi^\mu = \overline{\chi} \gamma^\mu (g_V^\chi + g_A^\chi \gamma_5 ) \chi \, .
\end{align}
Here, $g_V^f$ and $g_A^f$ are the vector and axial-vector couplings of SM fermions to the vector mediator $V_{\mu}$, while $g_V^\chi$ and $g_A^\chi$ define the vector and axial-vector couplings between the mediator $V_{\mu}$ and $\chi$. The Lagrangian contains both left-and right-handed currents, thus implicitly assuming the presence of an extended neutrino sector either containing sterile neutrinos or right-handed species. We will not go into the details of such an extended neutrino sector, simply assuming the presence of such left-and right-handed currents and dealing with their phenomenology. 

Following the general philosophy of simplified DM models~\cite{Abdallah:2015ter,DeSimone:2016fbz}, we write our effective theory after electroweak symmetry breaking (EWSB), assuming all the couplings to be independent. This raises questions about possible constraints coming from embedding such simplified models into a consistent UV-completion. The case of a $U(1)$ gauge extension has for example been studied in~\cite{Kahlhoefer:2015bea}, where it has been shown that, depending on the vector-axial nature of the couplings between the vector and the fermions (either DM or SM particles), large regions of parameter space may be excluded.\footnote{Even without considering the additional fermionic content which may be needed to make the model anomaly free.} In our case we allow the possibility of different couplings between the members of weak doublets, {\it  i.e.} terms with isospin breaking independent from the EWSB. As shown in~\cite{Fox:2011qd}, such a possibility arises, for example, at the level of $d=8$ operators, implying that the couplings are likely to be very small. Moreover, we expect a non-trivial contribution of the  isospin breaking sector to electroweak precision measurements, in particular to the $T$ parameter. Since the analysis is however highly model dependent, we will not pursue it here.

We briefly comment here on collider limits for the new neutral vector boson. For $m_V<209$ GeV, limits from LEP I~\cite{Abreu:1994ria}, analyzing the channel $e^+ e^- \to \mu^+ \mu^-$,  impose that its mixing with the Z boson has to be $\lsim 10^{-3}$,  implying the new gauge coupling to be $\lsim 10^{-2}$. This limit can be evaded in the case of a $U(1)$ gauge extension, if the new charges are not universal and the new boson does not couple (or couples very weakly) to muons or by  small $U(1)$ charges in extensions involving extra scalar fields (See, for instance, eq.(2.5) of ref.~\cite{Carena:2004xs}). For $m_V>209$ GeV, there are also limits from LEP II~\cite{Agashe:2014kda}, Tevatron~\cite{Carena:2004xs} and the LHC~\cite{Aad:2012hf,Chatrchyan:2012oaa,ATLAS:2012pu}. Since these limits depend on the fermion $U(1)$ charges, they can be either avoided or highly suppressed.

\subsection{Scalar mediator}
For a scalar mediator, we extend the SM by adding a Dirac fermion DM, $\chi$, with mass $m_\chi$ and a real scalar boson, $S$, with mass $m_S$. The relevant terms in the Lagrangian with the associated currents are
\begin{align}
\label{eq:simp_model_scal}
{\cal L}_{\rm sc} &= S\left(\sum_f g_S^f\, \overline{f}\, f   + g_S^\chi\,  \overline{\chi}\, \chi\right) - \displaystyle \frac{1}{2}\,m_S^2 S^2 \, .
\end{align}
The couplings $g_S^f $ and $g_S^\chi$ define the interaction between the scalar and the SM and the DM sectors, respectively.  Similar to the vector mediator interactions, the presence of an extended neutrino sector is assumed in the scalar mediator Lagrangian as well. Since in this work we will focus on the spin-independent cross-section at direct detection experiments, we consider only the possibility of a CP even real scalar mediator.

For a scalar singlet it is easier to imagine a (possibly partial) UV-completion. Take, for example, the case of a singlet scalar field $S$ added to the SM, which admits a quartic term $V \subset\lambda_{HS} |H|^2 S^2$ and takes the vacuum expectation value (VEV) $v_S$. Non local dimension 6 operators such as $ S^2 \overline{\ell}_L H e_R$ are generated (with $\ell_L$ and $e_R$ being the SM lepton doublets and singlets), which after spontaneous symmetry breaking and for energies below the Higgs boson mass $m_H$ produce the couplings of eq.~(\ref{eq:simp_model_scal}) as $g_S^f = y_f \sin\alpha$, where $\alpha = \frac{1}{2} \arctan(\lambda_{HS} v v_S/(m_S^2-m_H^2) )$ and $y_f$ is the fermion Yukawa coupling. The coupling with neutrinos can be arranged for example in the case of a neutrinophilic 2HDM~\cite{Gabriel:2006ns,Davidson:2009ha}. Typical values of $g_S^f$ can be inferred from the specific realization of the simplified model, but in the remainder of this paper, we will remain agnostic about realistic UV-completions in which the simplified models we consider can be embedded, focusing only on the information that can be extracted from CNSN. We stress however that, depending upon the explicit UV-completion, other constraints apply and must be taken into account to assess the viability of any model. For example, in generic BSM scenarios with a new mediator coupling to fermions, the dijet and dilepton analyses at the LHC put important bounds, as has been exemplified in~\cite{Chala:2015ama, Arcadi:2013qia}. In this paper we aim to concentrate only on the constraints arising from direct detection experiments.

\section{Scattering at direct detection experiments}
\label{sec:dd}
\subsection{Neutrino and dark matter Scattering}
\label{sec:direct_neutrino}
Let us now remind the reader about the basics of CNSN. In the SM, coherent neutrino scattering off nuclei is mediated by neutral currents. The recoil energy released by the neutrino scattering can be measured in the form of heat, light or phonons. The differential cross-section in terms of the nuclear recoil energy $E_R$ reads~\cite{Billard:2013qya}
\begin{align}
\label{eq:recoil_SM}
 \left. \frac{d\sigma^{\nu}}{dE_R} \right|_{\rm SM} &= ({\cal Q}_V^{\rm SM})^2 {\cal F}^2(E_R) \frac{G_F^2 m_N}{4\pi}   \left(1-\frac{m_N E_R}{2 E_\nu^2} \right) \, ,
\end{align}
with the SM coupling factor
\begin{align}
 {\cal Q}_V^{\rm SM} = N + (4 s_W^2 -1) Z\,.
\end{align}
Here, $N$ and $Z$ are the number of neutrons and protons in the target nucleus, respectively, $ {\cal F}(E_R)$ the nuclear form factor, $E_\nu$ the incident neutrino energy and $m_N$ the nucleus mass $G_F$ is the Fermi constant and $s_W=\sin\theta_W$ is the sine of the weak mixing angle. In addition, we use the nuclear form factor~\cite{Lewin:1995rx}
\begin{align}
{\cal F}(E_R)&=3\,\frac{j_1\left(q(E_R)r_N\right)}{q(E_R)r_N}\exp\left(-\frac{1}{2}[s \, q(E_R)]^2\right),
\end{align} 
 where $j_1(x)$ is a spherical Bessel function, $q(E_R)=\sqrt{2 m_n (N+Z) E_R}$ the momentum exchanged during the scattering, $m_n \simeq 932$~MeV the nucleon mass, $s\sim 0.9$ the nuclear skin thickness and $r_N\simeq 1.14 \, (Z+N)^{1/3}$ is the effective nuclear radius.

In the case of the vector model defined in eq.~(\ref{eq:simp_model_vec}), the differential cross-section gets modified by the additional $V$ exchange. The total cross-section should be calculated as a coherent sum of SM $Z$ and vector $V$ exchange, and reads
\begin{align}
\label{eq:recoil_vector}
{\displaystyle \left. \frac{d\sigma^{\nu}}{dE_R}\right|_{\rm V}} &= {\displaystyle {\cal G}_V \left. \frac{d\sigma^{\nu}}{dE_R}\right|_{\rm SM} \, , }
\end{align}
with
\begin{equation}
{\cal G}_V = 1 + \frac{4}{G_F^2} \left(\frac{{\cal Q}_V}{{\cal Q}_V^{\rm SM}}\right)^2 \frac{(g_V^\nu)^2 + (g_A^\nu)^2}{(q^2-m_V^2)^2} - \frac{2\sqrt{2}}{G_F} \frac{{\cal Q}_V}{{\cal Q}_V^{\rm SM}} \frac{g_V^\nu - g_A^\nu}{q^2-m_V^2} \, .
\end{equation}
Here, the coupling factor ${\cal Q}_V$ of the exotic vector boson exchange is given by \cite{Agrawal:2010fh}
\begin{align}
	{\cal Q}_V = (2Z+N) g_V^u + (2N+Z) g_V^d\,,
\end{align}
and $q^2 = - 2 \, m_N E_R$ is the square of the momentum transferred in the scattering process. To obtain eq.~(\ref{eq:recoil_vector}), we assumed that the neutrino production in the sun is basically unaffected by the presence of NP, in such a way that only LH neutrinos hit the target. As expected, if the new vector interacts only with RH neutrinos $g_V^\nu = g_A^\nu$, the NP contribution vanishes completely and no modification to the CNSN is present. On the other hand, when $g_V^\nu \neq g_A^\nu$, the interference term proportional to $g_V^\nu - g_A^\nu$ can give both constructive and destructive interference; in particular, remembering that  $q^2 - m_V^2 = - (2 \, m_N E_R + m_V^2)$ is always negative, we have constructive interference for $g_V^\nu < g_A^\nu$. For a detailed discussion of the interference effects at direct detection using effective theory formalism, see~\cite{Catena:2015qad}. As a last remark, let us notice that, due to the same Dirac structure of the SM and NP amplitudes, the correction to the differential cross-section amounts to an overall rescaling of the SM one.

For the simplified model with a scalar mediator defined in eq.~(\ref{eq:simp_model_scal}), the differential cross-section has a different form,
\begin{align}
\label{eq:recoil_scalar}
{\displaystyle\left. \frac{d\sigma^{\nu}}{d E_R} \right|_{\rm S}}  &= {\displaystyle\left. \frac{d\sigma^{\nu}}{d E_R} \right|_{\rm SM} + {\cal F}^2(E_R) \frac{{\cal G}_S^2 G_F^{2}}{4\pi} \, \frac{m_S^4 E_R m_N^2 }{E_\nu^2 (q^2-m_S^2)^2} \, ,} \quad \text{with} \quad
{\displaystyle{\cal G}_S }= {\displaystyle \frac{|g_S^\nu| {\cal Q}_S}{G_F \, m_S^2} \, .}
\end{align}
In this case, the modified differential cross-section is not simply a rescaling of the SM amplitude, but due to the different Dirac structure of the $S\bar{\nu}\nu$ vertex with respect to the SM vector interaction, it may in principle give rise to modification of the shape of the distribution of events as a function of the recoil energy. However, as we will see, for all practical purposes the impact of such modification is negligible.

Using the analysis presented in~\cite{Agrawal:2010fh}, the coupling factor for the scalar boson exchange is given by
\begin{align}
{\cal Q}_S &= Z m_n \left[\sum_{q=u,d,s} g_S^q \frac{f_{Tq}^p}{m_q}  +\frac{2}{27} \left( 1- \sum_{q=u,d,s}  f_{Tq}^p \right) \sum_{q=c,b,t} \frac{g_S^q}{m_q}  \right] \nonumber\\
  &+ N m_n  \left[\sum_{q=u,d,s} g_S^q \frac{f_{Tq}^n}{m_q}  +\frac{2}{27} \left( 1- \sum_{q=u,d,s}  f_{Tq}^n \right) \sum_{q=c,b,t} \frac{g_S^q}{m_q}  \right] \, .
\end{align}
The form factors $f_{Tq}^p, f_{Tq}^n$ capture the effective low energy coupling of a scalar mediator to a proton and neutron, respectively, for a quark flavor $q$.  For our numerical analysis we use 
$f_{Tu}^p = 0.0153$, $f_{Td}^p=0.0191$, $f_{Tu}^n = 0.011$, $f_{Td}^n = 0.0273$ and $f_{Ts}^{p,n} = 0.0447$, which are the values found in micrOMEGAs~\cite{Belanger:2014vza}. 
A more recent determination of some of these form factors can be found in Ref.~\cite{Alarcon:2011zs,Alarcon:2012nr}, we have used this estimation to 
determine the effect of the form factors on our final result (see Sec.~\ref{sec:modified_spectra}). 

A comment about the normalization of ${\cal G}_V$ and ${\cal G}_S$ is
in order. ${\cal G}_V = 1$ indicates recovery of purely SM
interactions with no additional contributions from exotic
interactions. For the scalar case, the situation is much
different. ${\cal G}_S$ includes ${\cal Q}_S$, a quantity dependent on
the target material. For LUX, ${\cal Q}_S \approx 1362 \,g_S^q$,
considering universal quark-mediator couplings, hence for
$|g_S^\nu|\sim 1$, $|g_S^q|\sim 1$, $m_S\sim 100$ GeV, natural values
of ${\cal G}_S$ are $\sim 10^4$. Turning now to DM, its scattering off
the nucleus can give rise to either spin-independent or spin-dependent
interactions. In our analysis we will consider only the
spin-independent scattering~\footnote{For the mediators we consider here, 
the spin-dependent cross-section is in fact velocity suppressed by $v^2$, see 
for instance \cite{Kumar:2013iva}.}, as the next generation experiments
sensitive to this interaction will also be sensitive to neutrino
scattering events. The spin-independent differential cross-section in
each of the two simplified models is given by
\begin{align}
 \left. \frac{d\sigma^{\chi}_{SI}}{dE_R} \right|_V &= 
 		{\cal F}^2(E_R) \frac{(g_V^\chi)^2 {\cal Q}_V^2}{4\pi}\frac{m_\chi m_N}{E_\chi (q^2-m_V^2)^2} \, , \\ 
 \left. \frac{d\sigma^{\chi}_{SI}}{dE_R} \right|_S &= 
 		{\cal F}^2(E_R) \frac{(g_S^\chi)^2 {\cal Q}_S^2}{4\pi} \frac{m_\chi m_N}{E_\chi (q^2-m_S^2)^2} \, ,
\end{align}
with the energy $E_\chi$ of the incident DM particle and all other variables as previously defined.

\subsection{Recoil events induced by DM and neutrino scattering}

Given the detector exposure, efficiency and target material, the above specified differential cross-sections can be converted into recoil event rates. 

We first look at the recoil event rate induced by neutrino scattering, where the differential recoil rate is given by
\begin{align}
\ds{\left.\frac{dR}{dE_R}\right|_\nu} &= \mathcal{N}\int_{E^{\nu}_{\rm min}}\ds{\frac{d\Phi}{dE_{\nu}}}\,\ds{\frac{d\sigma^{\nu}}{dE_R}}dE_{\nu}\, .
\end{align}
Here, $\mathcal{N}$ is the number of target nuclei per unit mass, $d\Phi/dE_{\nu}$ the incident neutrino flux and $E^{\nu}_{\rm min}=\sqrt{m_N\,E_R/2}$ is the minimum neutrino energy. $d\sigma^{\nu}/dE_R$ is the differential cross-section as computed in Eqs.~(\ref{eq:recoil_vector})-(\ref{eq:recoil_scalar}) for the vector and scalar  mediator models, respectively. For our numerical analysis, we use the neutrino fluxes from~\cite{Strigari:2009bq}. Integrating the recoil rate from the experimental threshold $E_{\rm th}$ up to $100$~keV~\cite{Billard:2013qya}, one obtains the number of neutrino events 
\begin{align}\label{eq:nu_events}
 {\rm{Ev}} ^\nu &= \int_{E_{\rm th}}\, \left.\frac{dR}{dE_R}\right|_\nu \, \varepsilon(E_R)\, dE_R\, ,
\end{align}
to be computed for either the scalar or the vector mediator models. Here, $E_\text{th}$ is the detector threshold energy and $\varepsilon(E_R)$ is the detector efficiency function. 

For the DM scattering off nuclei, the differential recoil rate also depends on astrophysical parameters such as the local DM density, the velocity distribution and it is given as
\begin{align}
\label{eq:rate_DM}
\left.\frac{dR}{dE_R}\right|_{\chi} &= \mathcal{N} \,\ds{\frac{\rho_0}{m_N\,m_{\chi}}} \int_{v_{\rm min}}v f(v)\frac{d\sigma^{\chi}_{SI}}{dE_R}d^3v\, ,
\end{align}
where $\rho_0=0.3$~GeV/$c^2$/cm$^3$ is the local DM density~\cite{Agrawal:2010fh},  $v$ is the magnitude of the DM velocity, $v_{\rm min}(E_R)$ is the  minimum DM speed required to cause a nuclear recoil with energy $E_R$ for an elastic collision and  $f(v)$ the DM   velocity distribution in the Earth's frame of reference.  This distribution is in principle modulated in time due to the Earth's motion around the Sun, but we ignore this effect here as it is not relevant for our purposes. If the detector has different target nuclides, one has to sum over all  their weighed contributions as, for instance, is done in  Ref.~\cite{DelNobile:2013gba}.

In what follows we will assume a Maxwell-Boltzmann distribution, given as
\begin{align}
f(v) = \left\{
\begin{array}{lcl}
\ds{\frac{1}{N_{\rm esc}\,(2\pi\,\sigma_v^2)^{3/2}}}\,\exp\Big[\ \ds{\frac{-(v + v_{\rm lab})^2}{2\sigma_v^2}}\Big] &~&  | v + v_{\rm lab}| < v_{\rm esc}, \\
 0 &~&   | v +   v_{\rm lab}| \geq v_{\rm esc},
\end{array} \right.
\end{align}
where $v_{\rm esc}=544$ km s$^{-1}$, $v_{\rm lab}=232$ km s$^{-1}$ and $N_{\rm esc}=0.9934$ is a normalization factor taken from~\cite{Agrawal:2010fh}.

In order to constrain only the DM-nucleon interaction cross-section $\sigma_{\chi n}$ at zero momentum transfer, which is independent of the type of experiment, it is customary to write eq.~(\ref{eq:rate_DM}) as
\begin{align}
\left.\frac{dR}{dE_R}\right|_{\chi} &= \mathcal{N}
\,\ds{\frac{\rho_0}{2\,\mu_n^2\,m_{\chi}}}\sigma_{\chi n} (Z+N)^2 {\cal
  F}(E_R)^2 \int_{\rm v_{min}}\frac{f(v)}{v}d^3v\, ,
\end{align}
where $\mu_n = m_n m_\chi / (m_n + m_\chi)$ is the reduced mass of the DM-nucleon system. For the cases we are considering (NP $= V$ or $S$), we have \cite{Agrawal:2010fh,Gelmini:2015zpa}
\begin{align}
\label{eq:DM_xsec_indep}
\sigma_{\chi n} &= \frac{\mu_n^2}{\mu_N^2} \frac{1}{(Z+N)^2} \int_0^{2 \mu_N^2 v^2/m_N} \left.\frac{d\sigma_{SI}(E_R=0)}{d E_R}\right|_{\rm NP} dE_R \nonumber \\
& =  \frac{(g^\chi_{NP})^2{\cal Q}_{\rm NP}^2}{m_{\rm NP}^4}\frac{\mu^2_n}{\pi (Z+N)^2}\, ,
\end{align}
where $\mu_N = m_N m_\chi / (m_N + m_\chi)$ is the reduced mass of the DM-nucleus system and we used that ${\cal F}(0)=1$. The number of DM events per ton-year can be obtained using an expression analogous to eq.~(\ref{eq:nu_events}), explicitly
\begin{align}
\label{eq:chi_events}
{\rm Ev} ^\chi &= \int_{E_{\rm th}}\, \left.\frac{dR}{dE_R}\right|_\chi \, \varepsilon(E_R)\, dE_R\,.
\end{align}
%
%

%
\subsection{Background free sensitivity in the presence of exotic neutrino interactions }

The presence of CNSN at direct detection experiments highlights the existence of a minimal DM - nucleon scattering cross-section below which CNSN events can not be avoided and in this sense the direct detection experiments no longer remain background free. This minimum cross-section is different for different experiments, depending on the detector threshold, exposure and target material. Using the definition given in eq.~(\ref{eq:DM_xsec_indep}), it is possible to represent the CNSN in the $(m_\chi, \sigma_{\chi n})$ plane introducing the so-called {\it one neutrino event contour line}. This line essentially defines the DM mass dependent threshold/exposure pairs that optimise the background-free sensitivity estimate at each mass while having a background of one neutrino event. The presence of additional mediators will modify this minimum cross-section with respect to the SM and hence, modify the maximum reach of an experiment. In this section, we show how the one neutrino event contour line changes due to the additional vector and scalar mediators considered in eq.~(\ref{eq:recoil_vector}) and~(\ref{eq:recoil_scalar}).

To compute the one neutrino event contour line we closely follow Ref.~\cite{Billard:2013qya}.  Considering, for instance, a fictitious Xe target experiment, we determine the exposure to detect a single neutrino event,
\begin{align}
	{\cal E}_\nu(E_{\rm th}) &= \frac{ \rm{Ev} ^\nu=1}{\int_{E_{\rm th}}\,  \left.\frac{dR}{dE_R}\right|_{\nu}\,dE_R} \, ,
\end{align}
as a function of energy thresholds in the range $10^{-4}\, {\rm keV} \leq E_{\rm th} \leq 10^{2}\, {\rm keV}$, varied in logarithmic steps. For each threshold we then compute the background-free exclusion limits, defined at 90\% C.L. as the curve in which we obtain 2.3 DM events for the computed exposure: 
\begin{align}\label{eq:xsec_nufloor}
	\sigma_{\chi n}^{1\nu} & = \frac{2.3}{{\cal E}_\nu(E_{\rm th})\,\int_{E_{\rm th}}\,  \left.\frac{dR}{dE_R}\right|_{\chi,\, \sigma_{\chi n}=1}\, dE_R}\, .
\end{align}
%

\begin{figure}[t!]
\centering
\includegraphics[width=\textwidth]{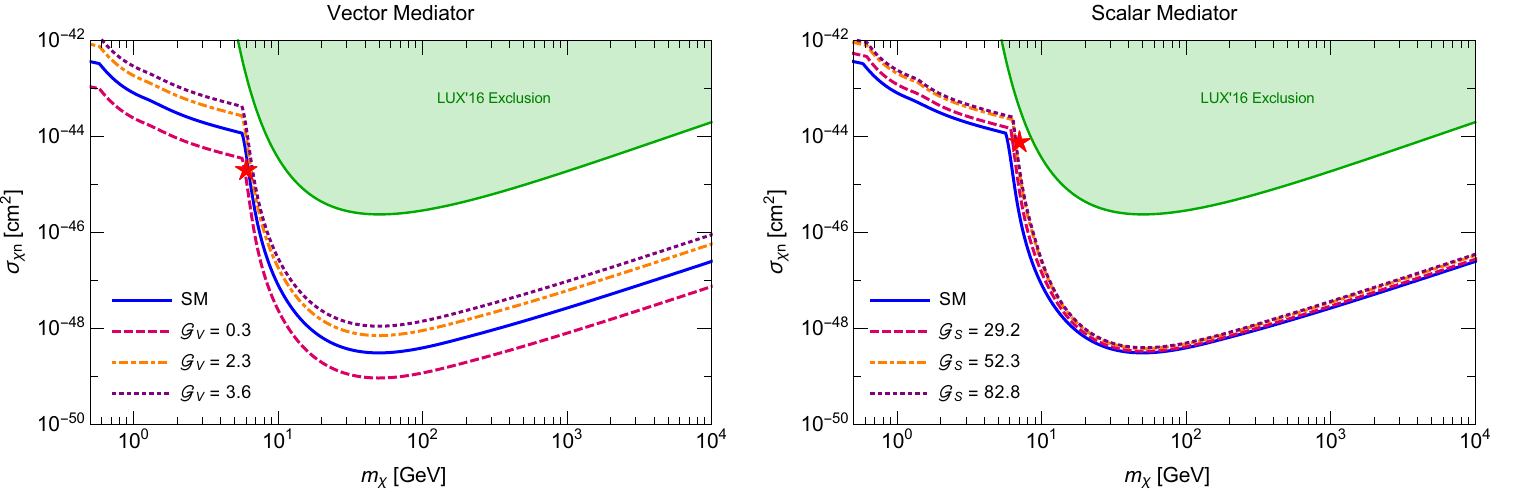}
\caption{One neutrino event contour lines for the two types of mediators, considering a Xe target detector. We show on the left (right) panel three examples for the vector (scalar) mediator. We also show the SM one neutrino event contour line (in blue) for comparison. The red star is a point for which we will show the energy spectrum. The green region is excluded by LUX at $90\%$ of C.L.~\cite{Akerib:2016vxi}.}
\label{fig:MNF-V}
\end{figure}
If we now take the lowest cross-section of all limits as a function of the DM mass, we obtain the one neutrino event contour line, corresponding to the best background-free sensitivity achievable for each DM mass for a one neutrino event exposure. Let us stress that the one neutrino event contour line, as defined in this section, is computed with a $100\%$ detector efficiency. The effect of a finite detector efficiency will be taken into account in Sec.~\ref{sec:modified_spectra} when we will compute how the new exotic neutrino interactions can affect the discovery potential of direct detection DM experiments. Comparing eq.~(\ref{eq:xsec_nufloor}) with Eqs.~(\ref{eq:rate_DM}) and (\ref{eq:DM_xsec_indep}), we see that the simplified models introduced in Sec.~\ref{sec:frame} can modify the one neutrino event contour line.  In fact, such modifications have been studied in specific models with light new physics {\it e.g.} in~\cite{Harnik:2012ni}. We show in Fig.~\ref{fig:MNF-V} some examples of a modified one neutrino event contour line for our models, fixing the values of the parameters $\mathcal{G}_V$ and $\mathcal{G}_S$ as specified in the legends. These parameters have been chosen to be still allowed by current data, see sections \ref{sec:limits} and \ref{sec:modified_spectra}. The left panel of the figure describes changes in the one neutrino event contour line in presence of a new vector mediator. As will be explained below, it is possible to have cancellation between SM and exotic neutrino interactions leading to a lowering of the contour line as shown for the case of ${\cal G}_V = 0.3$. It is also worth recollecting that ${\cal G}_V$ includes the SM contribution i.e. ${\cal G}_V = 1$ is the SM case. For the vector case the one neutrino event contour line is effectively a rescaling of the SM case. figure~\ref{fig:MNF-V} (right panel) on the other hand shows modification of the contour line for a scalar mediator. Note that unlike in the vector scenario, the factor ${\cal G}_S$ has a different normalization. No significant change in the one neutrino event contour line is expected in the scalar case.
\begin{table}[tb]
\centering 
\begin{tabular}{|c|c|c|}\hline 
Nucleus & $a^\nu_V $  & $a^{\nu}_S$ \\ \hline\hline  
 Xe & $1.0 \times 10^{-6}$ &  $4.5 \times 10^{-7}$ \\
 Ge & $9.4 \times 10^{-7}$ & $4.2 \times 10^{-7}$ \\
 Ca & $6.6 \times 10^{-7}$ & $3.7 \times 10^{-7}$ \\
 W  & $1.1 \times 10^{-6}$ & $4.5 \times 10^{-7}$ \\
 O  & $6.6 \times 10^{-7}$ & $3.7 \times 10^{-7}$ \\
\hline 
\end{tabular}
\caption{\label{tab:limits} Values of the coefficients $a^\nu_V$ and $a^\nu_S$ for various target nuclei, corresponding to strongest reduction of the CNSN cross session according  to Eqs.~(\ref{eq:vlim}) and (\ref{eq:slim}).}
\end{table}

There are a few remarks we should make here. First, it is possible, in the context of the vector mediator model, to cancel the SM contribution to CNSN and completely eliminate the neutrino background. For mediator masses heavy enough to neglect the $q^2$ dependence of the cross-sections, this happens when, c.f. with eq.~(\ref{eq:recoil_vector}), 
\begin{align}
\label{eq:vlim}
g^{\nu}_V  - g^{\nu}_A & = \frac{{\cal Q}_V^{\rm SM}}{{\cal Q}_V}\frac{G_F m_V^2}{\sqrt{2}}  = \frac{a^\nu_V}{g_V^q} \left( \frac{m_V}{\rm GeV}\right)^2 \,,
\end{align}
where for the last equality we assume $g_V^{u}=g_V^{d}=g_V^{q}$ and $a^\nu_V$ is a numerical value that depends only on the target nucleus. We show in table~\ref{tab:limits}  the values of $a^\nu_V$ for various nuclei.

Second, in the case of the scalar scenario, it is possible to compensate for only part of the SM contribution to the CNSN. Inspecting Eqs.~(\ref{eq:recoil_SM}) and~(\ref{eq:recoil_scalar}) we see that the positive scalar contribution can at most cancel the negative SM term depending on $E_R/E_\nu^2$, resulting in an effective increase of the cross-section. This is accomplished for 
\begin{align}
\label{eq:slim}
g^\nu_S &= \frac{{\cal Q}_V^{\rm SM}\,}{{\cal Q}_S} \frac{G_F m_S^2}{\sqrt{2}}= \frac{a^\nu_S}{g_S^q}\left( \frac{m_S}{\rm GeV}\right)^2 \,,
\end{align}
where again $a^\nu_S$ is a numerical value that depends only on the target nucleus. Its value for different nuclei are shown in table~\ref{tab:limits}. We show in the right panel of figure~\ref{fig:MNF-V} an example of this situation, orange line, ${\cal G}_S = 52.3$, corresponding to the case of $g_S^q = 1$ and $m_S = 100$ GeV.

Finally, we should note that the one neutrino event contour line only gives us a preliminary estimate of the minimum cross-sections that can be reached by a DM direct detection experiment. It is worth recalling that this estimate is a background-free sensitivity. Interactions modifying both neutrino and DM sector physics will lead to a non-standard neutrino CNSN background which should be taken into account. Furthermore, the compatibility of the observed number of events should be tested against the sum of neutrino and DM events. In this spirit, to answer the question {\em what is the DM discovery potential of an experiment?} one has to compute the real {\em neutrino floor}. This will be done in Sec.~\ref{sec:modified_spectra}, which will include a more careful statistical analysis taking into account background fluctuations and the experimental efficiency.

\section{Current and future limits on DM-neutrino interactions}
\label{sec:limits}

When new physics interacts with the DM and neutrino sector, the limits from direct detection experiments become sensitive to the sum of DM and neutrino scattering events. A natural question to ask is the capacity of current experiments to constrain this sum. The aim of this section is to assess these constraints and derive sensitivities for the next generation of direct detection experiments. For the analysis of the current limits we consider the results of the Large Underground Xenon (LUX)~\cite{Akerib:2016vxi} experiment. This choice is based on the fact that this experiment is at present the most sensitive one probing the $m_\chi> 5$ GeV region on which we focus.  On the other hand, for the future perspectives we will consider two Xe target based detectors: the one proposed by the LUX-ZonEd Proportional scintillation in LIquid Noble gases (LUX-ZEPLIN)  Collaboration~\cite{Akerib:2015cja} and the one proposed by the DARk matter WImp search with liquid xenoN (DARWIN)  Collaboration~\cite{Aalbers:2016jon}.\\

\noindent{\bf Current bounds}. LUX is an experiment searching for WIMPs through a dual phase Xe time projection chamber. We will consider its results after a $3.35 \times 10^4$ kg-days run presented in 2016~\cite{Akerib:2016vxi}, performed with an energy threshold of $1.1$ keV. We also use the efficiency function $\varepsilon(E_R)$ reported in the same work.

In order to assess the constraining power of current LUX results for the two models presented in Eqs.~(\ref{eq:simp_model_vec})-(\ref{eq:simp_model_scal}), we compute the total number of nuclear recoil events expected at each detector as
\begin{align}
{\rm Ev}^{\rm total} & = {\rm Ev}^{\chi} + {\rm Ev}^{\nu}.
\end{align}
Using this total number of events, we compute a likelihood function constructed from a Poisson distribution in order to use their data to limit the parameters of our models,
\begin{align}
\mathcal{L}(\hat{\theta}|N) & = P(\hat{\theta}|N)=\frac{(b+\mu(\hat{\theta}))^Ne^{-(b+\mu(\hat{\theta}))}}{N!}\, ,
\end{align}
where $\hat{\theta}$ indicates the set of parameters of each model, $N$ the observed number of events, $b$ the expected background and $\mu(\hat{\theta})$ is the total number of events ${\rm Ev}^{\rm total}$.  According to~\cite{Dent:2016iht} we use $N=2$ for the number of observed events and $b=1.9$ for the estimated background. Maximizing the likelihood function we can obtain limits for the different planes of the parameter space.
\begin{figure}[t!]
\centering
\includegraphics[width=\textwidth]{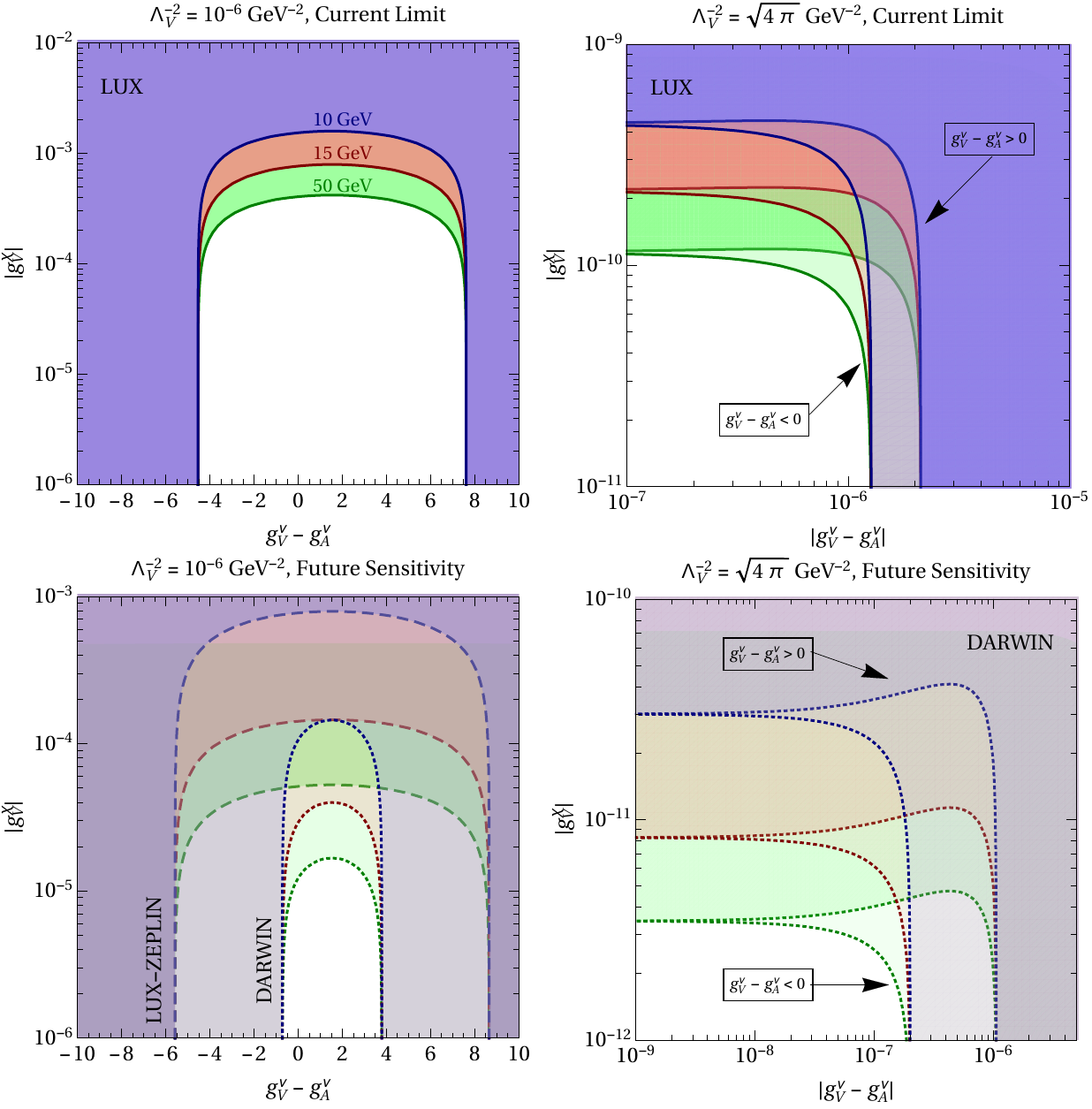} 
\caption{Current limits (top panels) and future sensitivity (bottom panels) on the parameters of the vector model. The coloured region can be excluded at 90\% C.L. by current LUX data~\cite{Akerib:2015rjg} (continuous lines) and by the future LUX-ZEPLIN~\cite{Akerib:2015cja} (dashed lines) and DARWIN~\cite{Aalbers:2016jon} experiments (dotted lines). The plots are for  $m_\chi = 10$ GeV  (violet), 15 GeV (red) and 50 GeV (green) for two different cases: $\Lambda^{-2}_V=10^{-6}$ GeV$^{-2}$ (left) and $\Lambda^{-2}_V=\sqrt{4\pi}$ GeV$^{-2}$ (right). For simplicity, in the latter case we only show the DARWIN future sensitivity, since the LUX-ZEPLIN results are qualitatively similar but a factor of $\sim$~4-10 less sensitive.}
\label{fig:V_lim}
\end{figure}

\begin{figure}[t!]
\centering
\includegraphics[width=\textwidth]{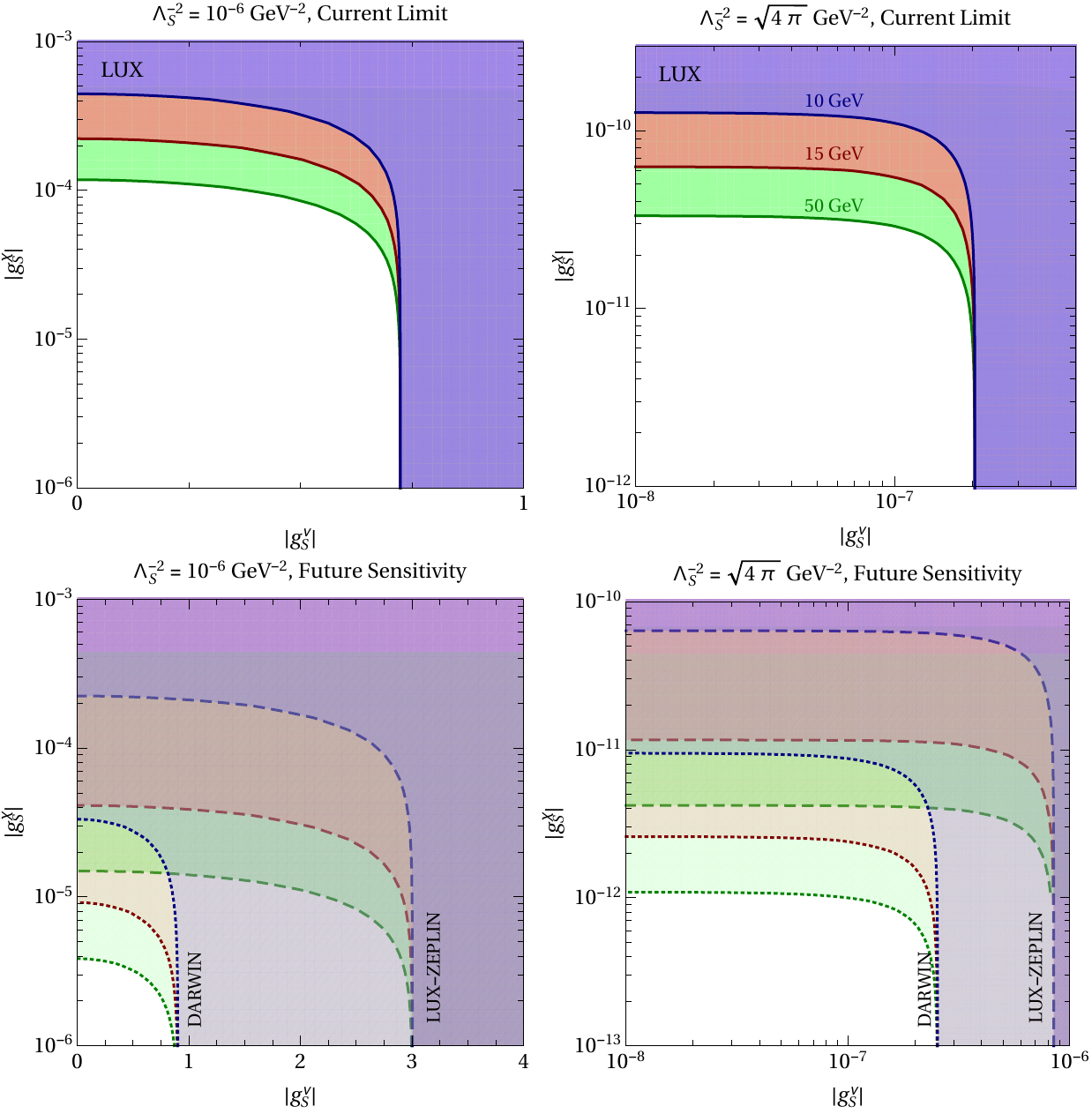} 
\caption{Current limits (top panels) and future sensitivity (bottom panels) on the parameters of the scalar model. The coloured region can be excluded at 90\% C.L. by current LUX data~\cite{Akerib:2015rjg} (continuous lines) and by the future LUX-ZEPLIN~\cite{Akerib:2015cja} (dashed lines) and DARWIN~\cite{Aalbers:2016jon} experiments (dotted lines). The plots are for  $m_\chi = 10$ GeV  (violet), 15 GeV (red) and 50 GeV (green) for two different cases: $\Lambda^{-2}_S=10^{-6}$ GeV$^{-2}$ (left) and $\Lambda^{-2}_S=\sqrt{4\pi}$ GeV$^{-2}$ (right).}
\label{fig:S_lim}
\end{figure}
In the case of the vector model, we performed a scan of the parameter space in the ranges
\begin{equation}
\label{eq:parameters_V}
  0\le \vert g_V^\nu - g_A^\nu \vert \le 10, \qquad
  0 \le \vert g_V^\chi\vert \le 1\, ,
\end{equation}
while we always choose $g_A^\chi=0$ in order to avoid spin-dependent limits. We show our limits in figure~\ref{fig:V_lim} for $\displaystyle \Lambda^{-2}_V \equiv g^q_{V}/m_V^2=10^{-6}$ GeV$^{-2}$~\footnote{For $g^q_V = 10^{-2}, 10^{-1}, 0.25, 0.5$ and 1, this corresponds, respectively, to $m_V \sim 100$ GeV, 315 GeV, 500 GeV, 710 GeV and 10$^{3}$ GeV.} and $\displaystyle\Lambda^{-2}_V = \sqrt{4\pi}$ GeV$^{-2}$ (right), which corresponds to a light mediator with $m_V = 1$ GeV and a coupling at the perturbativity limit. In each case, we show the results for three values of the DM mass, $m_\chi = 10$ GeV (violet), 15 GeV (red) and 50 GeV (green).  We see that we can clearly distinguish two regions: for $\Lambda^{-2}_V =10^{-6}$ GeV$^{-2}$, when $\vert g^\nu_V-g^\nu_A \vert \lesssim 3-4$, the DM contribution is the dominant one (in particular, as $\vert g^\nu_V - g^\nu_A \vert \to 0$ the contribution to the neutrino floor is at most the SM one), and sets $\vert g^\chi_V \vert \lsim 2 \times 10^{-3}$ ($\lsim 4 \times 10^{-4}$ ) for $m_\chi = 10 \,(50)$ GeV. On the other hand, for larger values of $\vert g^\nu_V-g^\nu_A \vert$, the number of neutrino events rapidly becomes dominant and no bound on the DM-mediator coupling can be set. For the extreme value $\Lambda^{-2}_V =\sqrt{4\pi}$ GeV$^{-2}$, one can set the limits $\vert g^\chi_V \vert \lsim 4.3 \times 10^{-10}$ ($\lsim 1.2 \times 10^{-10}$ ) for $m_\chi = 10 \,(50)$ GeV and $\vert g_V^\nu - g_A^\nu \vert \lsim$ few $10^{-6}$. 

Inspection of figure~\ref{fig:V_lim} shows two peculiar features: an asymmetry between the bounds on positive and negative values of $g_V^\nu-g^\nu_A$, and the independence of these limits on the DM mass. We see from eq.~(\ref{eq:recoil_vector}) that the asymmetry can be explained from the dependence of the interference term on the sign of $g_V^\nu-g^\nu_A$. Such interference is positive for $g_V^\nu-g^\nu_A<0$, explaining why the bounds on negative $g_V^\nu-g^\nu_A$ are stronger. As for the independence of the $g_V^\nu-g^\nu_A$ bounds from the DM mass, this can be understood from the fact that when $g_V^\chi$ becomes sufficiently small we effectively reach the $g_V^\chi\to 0$ limit in which the DM mass is
not relevant.

Turning to the bounds that the current LUX results impose on the parameter space of the scalar model, we varied the parameters in the ranges 
\begin{equation}
\label{eq:parameters_S}
	0 \le |g_S^\nu| \le 5   \qquad  0 \le |g_S^\chi| \le 1\, .
\end{equation}
Our results are presented in figure~\ref{fig:S_lim}. On the top left (right) panel, fixing $\displaystyle \Lambda^{-2}_S \equiv g^q_{S}/m_S^2=10^{-6}$ GeV$^{-2}$ ($\Lambda^{-2}_S =\sqrt{4\pi}$ GeV$^{-2}$), for $m_\chi = 10$ GeV (violet), 15 GeV (red) and 50 GeV (green). From these plots we see that LUX can limit $\vert g^\chi_S \vert \lsim 4.5 \times 10^{-4}$ ($\lsim 1 \times 10^{-4}$ ) for $m_\chi = 10 \,(50)$ GeV if $\vert g^\nu_S \vert < 0.5$, when $\Lambda^{-2}_S=10^{-6}$ GeV$^{-2}$. For the limiting case $\Lambda^{-2}_S=\sqrt{4\pi}$ GeV$^{-2}$, we get the bound $\vert g^\chi_S \vert \lsim 1.3 \times 10^{-10}$ ($\lsim 3.2 \times 10^{-11}$ ) for $m_\chi = 10\,(50)$ GeV if $\vert g^\nu_S \vert < 10^{-7}$.  

As $g^\nu_S \to 0$, the contribution to the neutrino floor tends to the SM one, except for a particular value of $g^\nu_S\, g^q_S$, as discussed at the end of the previous section. In the opposite limit, {\it i.e.} where the neutrino floor dominates, $g^\chi_S \to 0$, the current limit is $\vert g^\nu_S \vert \lsim 0.7$ ($\vert g^\nu_S \vert \lsim$ few $10^{-7}$) for $\Lambda^{-2}_S =10^{-6}$ ($=\sqrt{4\pi}$) GeV$^{-2}$. As in the vector case, we see that this bound does not depend on the DM matter mass, for the same reasons explained above.\\

\noindent{\bf Future sensitivity.} To assess the future projected LUX-ZEPLIN sensitivity, we will assume an energy threshold of $6$ keV, a maximum recoil energy of $30$ keV and a future exposure of $15.34$ t-years~\cite{Akerib:2015cja}. According to the same reference, we use a $50\%$ efficiency for the nuclear recoil.  For DARWIN, we will consider an aggressive 200 t-years exposure, no finite energy resolution but a 30\% acceptance for nuclear-recoil events in the energy range of 5--35 keV~\cite{Aalbers:2016jon}.

Let us now discuss the bounds that can be imposed on the parameter space of our models in case the future experiments LUX-ZEPLIN and DARWIN will not detect any signal. We scan the parameter space over the ranges of Eqs.~(\ref{eq:parameters_V})
and~(\ref{eq:parameters_S}), obtaining the exclusion at $90\%$ C.L. The results are presented in the bottom panels of figure~\ref{fig:V_lim} (figure~\ref{fig:S_lim}) for the vector (scalar) model.

In the region in which the DM events dominate, we see that LUX-ZEPLIN will be able to improve the bound on $\vert g_V^\chi \vert$ and $\vert g_S^\chi \vert$ by a factor between $2$ and $10$ depending on the DM mass, while another order of magnitude improvement can typically be reached with DARWIN. However, we also see that, somehow contrary to expectations, the bounds on the neutrino couplings are expected to be less stringent than the present ones. While the effect is not particularly relevant in the vector case, we can see that in the scalar case the LUX-ZEPLIN sensitivity is expected to be about a factor of $4$ worse than the current LUX limit. This is due to the higher threshold of the experiment, that limits the number of measurable solar neutrino events. As such, a larger $\vert g_S^\chi \vert$ is needed to produce a sufficiently large number of events, diminishing the constraining power of LUX-ZEPLIN. While in principle this is also true  for the DARWIN experiment, the effect is compensated by the aggressive expected exposure.

\section{Sensitivity to DM-nucleon scattering in presence of exotic neutrino interactions}
\label{sec:modified_spectra}

In section~\ref{sec:dd}, we computed the background-free sensitivity of direct detection experiments in presence of exotic neutrino interactions. However, what is the {\em true} $3\sigma$ discovery potential given the exotic neutrino interactions background remains unanswered. In this section, we perform a detailed statistical analysis, taking into account the estimated background and observed number of events and comparing these against the DM and neutrino interaction via a profile likelihood analysis.

To assess the DM discovery potential of an experiment we calculate, as in Ref.~\cite{Billard:2011zj}, the minimum value of the scattering cross-section $\sigma_{\chi n}$ as a function of $m_\chi$ that can be probed by an experiment. This defines  a discovery limiting curve that is the true {\em neutrino floor} of the experiment. Above this curve the  experiment has a 90\% probability of observing a 3$\sigma$ DM detection. This is done by defining a binned likelihood function~\cite{Ruppin:2014bra,O'Hare:2016ows}
\begin{align}
\label{eq:likelihood}
{\cal L}(\sigma_{\chi n},m_\chi,\phi_\nu,\mathbf{\Theta}) &= \prod_{i=1}^{n_{\rm b}} {\cal P} 
\left( {\rm Ev}^{\rm obs}_i \vert {\rm Ev}^\chi_i + \sum_{j=1}^{n_\nu} {\rm Ev}^{\nu}_i(\phi_\nu^j);\mathbf{\Theta} \right) \times  \prod_{j=1}^{n_\nu} {\cal L}(\phi_\nu^j) \, ,
\end{align}
where we have a product of Poisson probability distribution functions
(${\cal P}$) for each bin $i$ ($n_{\rm b}=100$), multiplied by
gaussian likelihood functions parametrizing the uncertainties on each
neutrino flux normalization, ${\cal L}(\phi_\nu^j)$
\cite{O'Hare:2016ows}. The neutrino (${\rm Ev}^\nu$) and DM (${\rm Ev}^\chi$) number of events were 
computed according to Eqs.~(\ref{eq:nu_events}) and (\ref{eq:chi_events}), 
respectively. For each neutrino component $j=1,\ldots,
n_\nu$, the individual neutrino fluxes from solar and atmospheric
neutrinos are denoted by $\phi_\nu^j$, while $\mathbf{\Theta}$ is a
collection of the extra parameters ($g_{V,S}^{q}$, $g_{V,S}^\nu$,
etc.) to be taken into account in the model under consideration. Since
we will introduce the discovery limit in the DM cross-section, note
that we will keep the DM-mediator coupling $g_{V,S}^\chi$ free. For
this study, we considered only the contribution of the $^8$B and $hep$
solar and atmospheric neutrinos, due to the thresholds of the
considered experiments. For a fixed DM mass, we can use
eq.~(\ref{eq:likelihood}) to test the neutrino-only hypothesis $H_0$
against the neutrino+DM hypothesis $H_1$ constructing the ratio
\begin{equation}
\lambda(0) = \frac{{\cal L}(\sigma_{\chi n}=0,\hat{\hat{\phi}}_\nu,\mathbf{\Theta})}{{\cal L}
(\hat{\sigma}_{\chi n},\hat{\phi}_\nu,\mathbf{\Theta})}\, ,
\end{equation}
where $\hat{\phi}_\nu$ and $\hat{\sigma}_{\chi n}$ are the values of the fluxes and DM cross-section that maximize 
the likehood function ${\cal L} (\hat{\sigma}_{\chi n},\hat{\phi}_\nu,\mathbf{\Theta})$, while $\hat{\hat{\phi}}_\nu$ is used to maximize ${\cal L}(\sigma_{\chi n}=0,\hat{\hat{\phi}}_\nu,\mathbf{\Theta})$. For each mass $m_\chi$ and cross-section $\sigma_{\chi n}$ we build a probability density function $p(Z\vert H_0)$ of the test statistics under $H_0$, the neutrino only hypothesis. This is performed by constructing an ensemble of 500 simulated experiments, determining for each one the significance $Z=\sqrt{-2 \ln \lambda(0)}$ ~\cite{Billard:2011zj,Ruppin:2014bra,O'Hare:2016ows}. Finally, we compute the significance that can be achieved 90\% of the times, $Z_{90}$, given by 
\begin{align}
\int_0^{Z_{90}} p(Z\vert H_0) \, dZ &=0.90 \, .
\end{align}
Therefore, the minimum value for the cross-section for which the experiment has 90\% probability of 
making a 3$\sigma$ DM discovery is defined as the value of $\sigma_{\chi n}$ that corresponds to $Z_{90}=3$.

\begin{figure}[t!]
\centering
\includegraphics[width=0.98\linewidth]{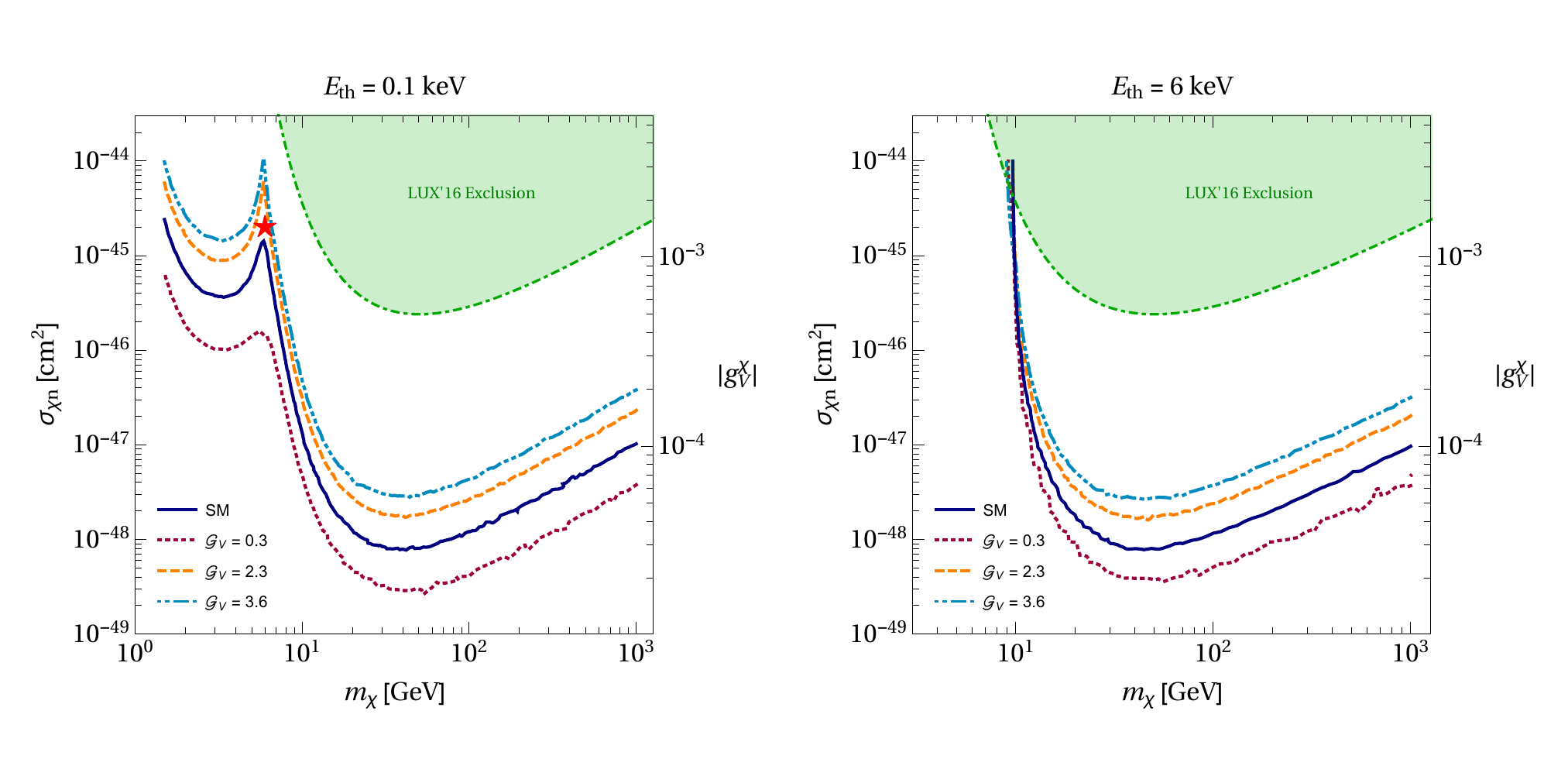}
\caption{\label{fig:nufloor-v} Neutrino floor for the vector mediator case in the plane ($m_\chi, \sigma_{\chi n}$) and  ($m_\chi, \vert g^\chi_V \vert$). The results are for the LUX-ZEPLIN experiment with two different energy thresholds: a very low one, $E_{\rm th}$ =0.1 keV (left), and the nominal threshold used in the experiment, $E_{\rm th}$ = 6 keV (right). The SM neutrino floor (dark blue) is shown, along with the most extreme case still allowed for the vector model (${\cal G}_V=3.6$, light blue), an intermediate case (${\cal G}_V=2.3$, orange), as well as a case where the neutrino floor can be smaller than the SM one (${\cal G}_V=0.3$, red). The axis corresponding to the $\vert g^\chi_V \vert$ coupling was obtained 
considering $\Lambda_V^{-2}=10^{-6}$ GeV$^2$.}
\end{figure}

In figure~\ref{fig:nufloor-v} we can see the neutrino floor considering
only the SM contribution to the CNSN (dark blue) as well as the result
for some illustrative cases, in the vector mediator scenario, for the
LUX-ZEPLIN experiment with two different energy thresholds.  The case
${\cal G}_V=3.6$ (light blue) can be considered an extreme case,
corresponding to the current limit on $\vert g_V^\nu - g_A^\nu \vert$
($\lsim 10^{-6}$) for $\Lambda_V^{-2}= \sqrt{4 \pi}$ GeV$^{-2}$. Above
this curve a 3$\sigma$ DM discovery can be achieved by the experiment,
while below this curve it is difficult to
discriminate between a DM signal and a non-standard (vector mediated)
contribution to the neutrino floor. We also show the case ${\cal
  G}_V=0.3$ (red), where the new vector contribution comes with the
opposite sign to the SM one, so it actually cancels some of the
standard signal. On the other hand, in the case corresponding to the
threshold $E_{th}=0.1$ keV, we find the same phenomenon noticed in the
literature: close to a DM mass of 6 GeV, the discovery limit is
substantially worsened because of the similarity of the spectra of
$^8$B neutrinos and the WIMP, see, for instance
\cite{Billard:2011zj}. However, the minimum cross-section that can be
probed is different for each parameter $\mathcal{G}_V$, due to the
contribution of the vector mediator. For the case of $E_{th}=6$ keV,
we see that the vector mediator decreases or increases the discovery
limit according to the value of ${\cal G}_V$.

\begin{figure}[t!]
\centering
\includegraphics[width=0.98\linewidth]{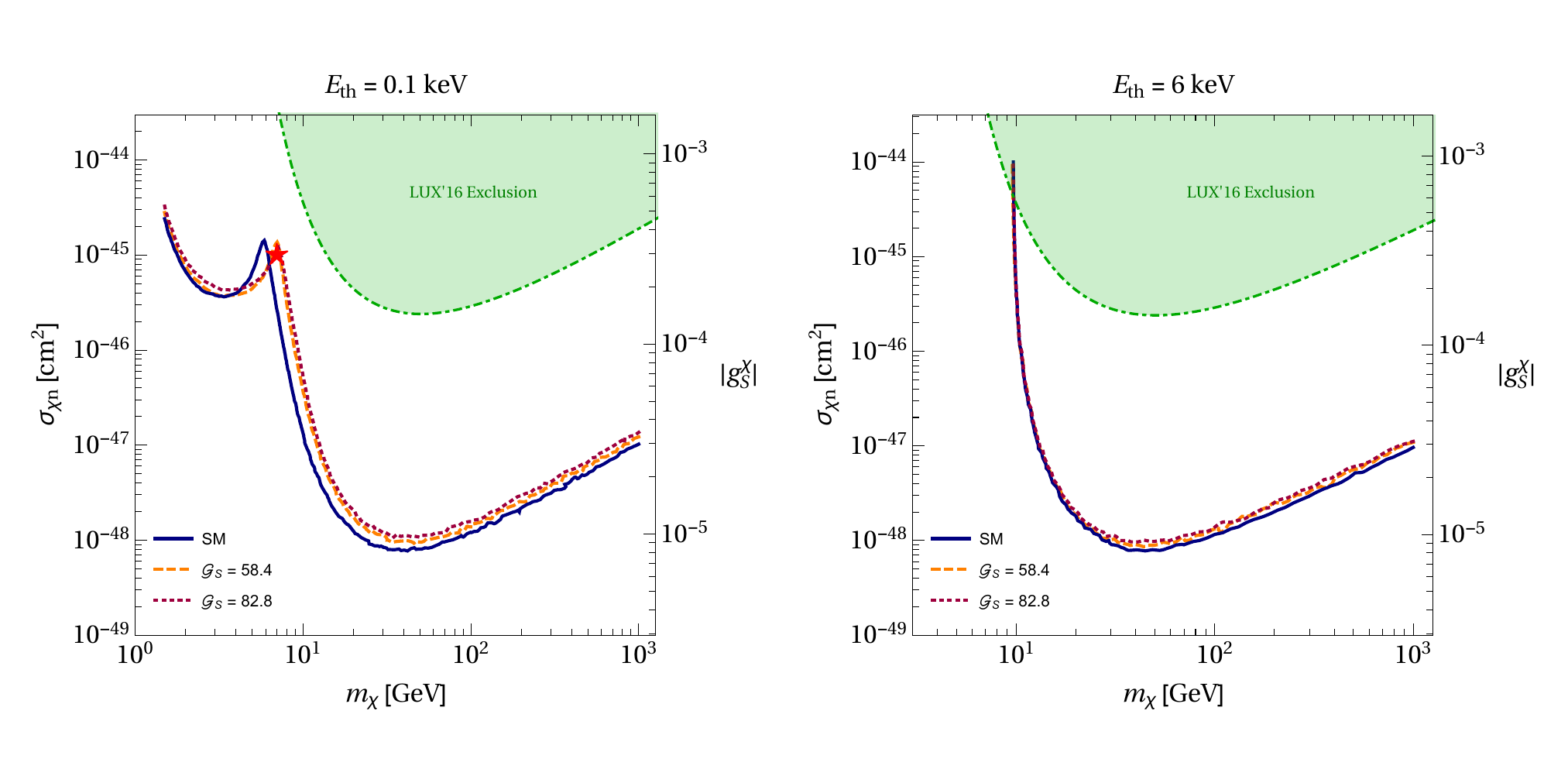}
\caption{\label{fig:nufloor-s} 
As figure~\ref{fig:nufloor-v}, but for the scalar mediator case. The SM neutrino floor (dark blue) is shown, along with two cases still allowed by the scalar model (${\cal G}_S = 58.4$, orange; ${\cal G}_S = 82.8$, red).}
\end{figure}

In figure~\ref{fig:nufloor-s} we can see the neutrino floor considering
only the SM contribution to the CNSN (dark blue) as well as the result
for some illustrative cases, in the scalar mediator scenario, for the
LUX-ZEPLIN experiment with two different energy thresholds.  Here the
case ${\cal G}_S= 82.8$ (orange) can be considered an extreme case,
since this corresponds to the current limit on $\vert g_S^\nu \vert$
($\lsim 2\times 10^{-7}$) for $\Lambda_S^{-2}= \sqrt{4 \pi}$
GeV$^{-2}$.  In the case of the lower threshold, we see that the point
where the discovery limit is highly affected due to the $^8$B
neutrinos is displaced close to a mass of 7 GeV. 
This shift of the distribution is provoked by the extra factor 
that appears in the scalar case with respect to the SM (see 
eq.~(\ref{eq:recoil_scalar})).
For the case of $E_{\rm th}=6$ keV, the scalar mediator
does not modify significantly the discovery limit. Therefore, we see
that contrary to the vector case, the scalar contribution does not
affect very much the discovery reach of the experiment as compared to
the one limited by the standard CNSN.

\begin{figure}[t!]
\centering
\includegraphics[width=0.8\linewidth]{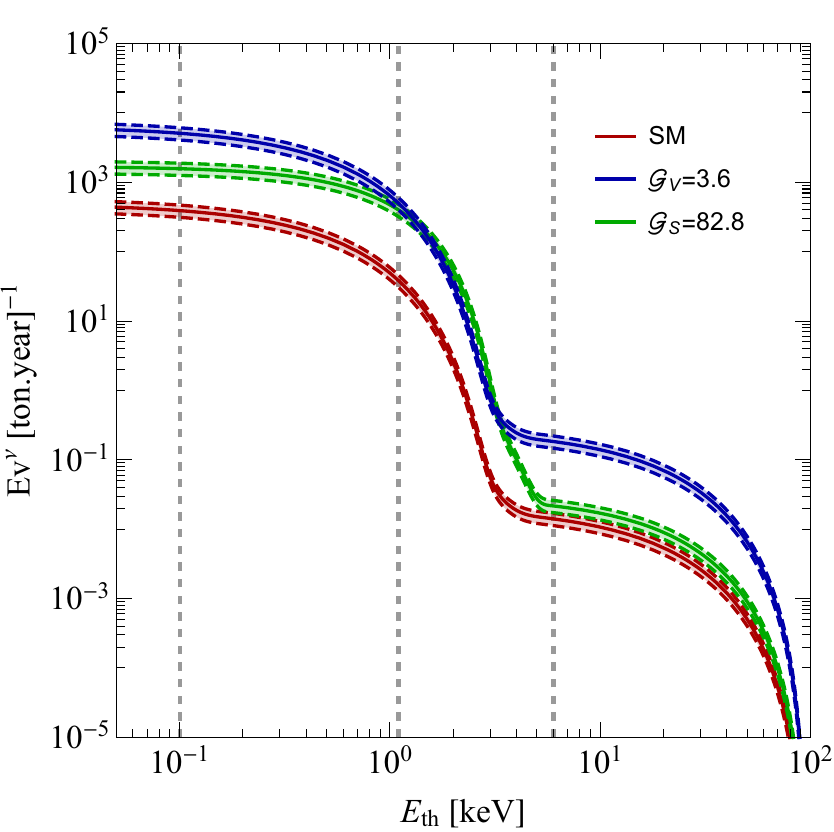}
\caption{Number of CNSN events per ton year for LUX-ZEPLIN as a functions of the energy threshold. In red we show the predictions for the 
SM and in blue (green)  for the vector (scalar) model with ${\cal G}_V=3.6$ (${\cal G}_S=82.8$).
The thickness of the curves represent a variation on the detector efficiency of 50\% $\pm$ 10\%.
}
\label{fig:threshold}
\end{figure}

In figure~\ref{fig:threshold} we show the behavior of the number of CNSN
events as a function of the energy threshold of the detector and for a
detector efficiency varying from 40\% to 60\%.  From this we see that
for ${\cal G}_V=3.6$ and ${\cal G}_S=82.8$, values that saturate the
current limit of the LUX experiment for the vector and scalar mediator
models, the number of neutrino events for $E_{\rm th}\sim 1$ keV are
basically the same and both about 10 times larger than the SM
contribution.  However, for the choices $E_{\rm th}\sim 0.1$ keV
(lower threshold) and $E_{\rm th}\sim 6$ keV (higher threshold) used
in Figs.~\ref{fig:nufloor-v} and \ref{fig:nufloor-s}, the number of
CNSN events for the vector model is about 4 times larger than that for
the scalar model, explaining the difference in sensitivity for the
vector and scalar models at those thresholds. We see again that the SM
and the vector mediator model number of CNSN events differ simply by a
scale factor, independent of the energy threshold, as expected from
eq.~(\ref{eq:recoil_vector}).  On the other hand, for the scalar case
there is a non-trivial behavior with respect to the SM due to the
extra term in the cross-section 
that depends on $E_R m_N/E_\nu^2$ (see eq.~(\ref{eq:recoil_scalar})), thus on $E_{\rm th}$.  For the lower
threshold low energy $^8$B neutrinos become accessible. However, the difference
between the SM and the scalar mediator cross-sections diminishes more
with lower $E_{\rm th}$ than it increases with lower $E_\nu$ so the number
of CNSN events differs only by a factor $\sim$ 3.  For the higher
threshold only atmospheric neutrinos are available, both SM and scalar
contributions are expected to be of the same order as the extra scalar
contribution is suppressed by $E_\nu^{-2}$.  We also see that a
detector efficiency between 40\% to 60\% does not affect the above
discussion and consequently we do not expect the neutrino floors we
have calculated in this section to be very different had we chosen to
use in our computation 40\% or 60\% efficiency instead of the 50\% we
have used.

\begin{figure}[t!]
\centering
\includegraphics[width=\linewidth]{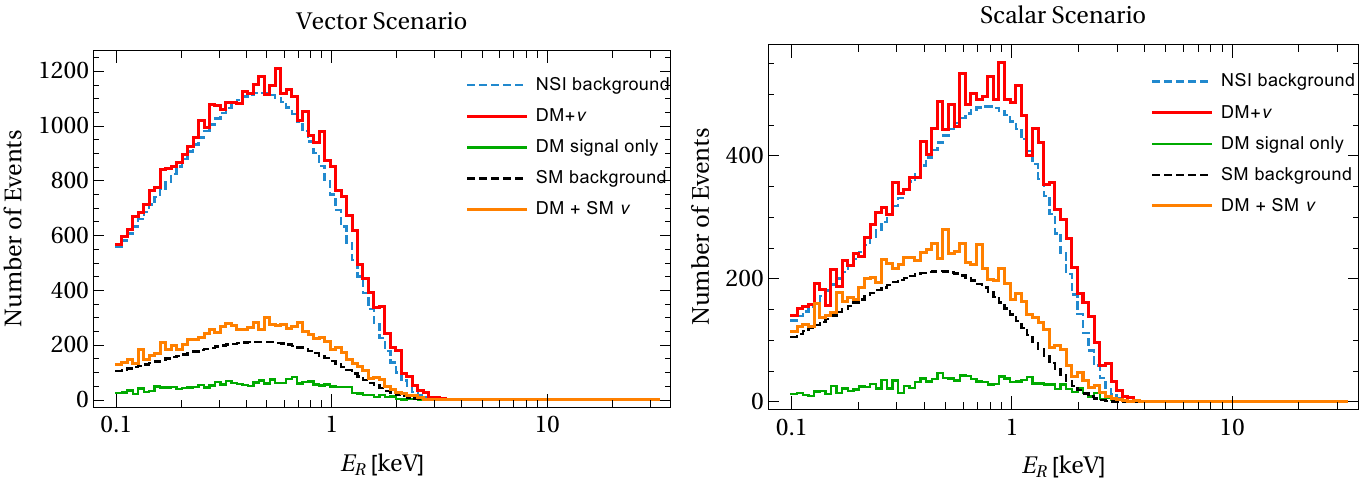}
\caption{Recoil spectrum in the vector (left) and scalar (right) case for the parameter point corresponding to the red star in figure~\ref{fig:nufloor-v} and figure~\ref{fig:nufloor-s}, respectively. The different contributions are shown separately: DM only (green), standard CNSN (black), non-standard CNSN (blue) as well as the combined spectrum (red).}
\label{fig:mod_spectra1}
\end{figure}

We have also performed an estimation of the effect of the uncertainty on the 
form factors $f_{Tq}^{p,n}$ on the results of our calculation and concluded that 
they can affect the neutrino floor by $\sim$ 30\%.

To exemplify the difficulty in discriminating between an energy spectrum produced by DM collisions from the modified neutrino floor, in the two cases studied in this paper, we show in figure~\ref{fig:mod_spectra1} examples of the energy spectrum for the points corresponding to the red stars in figure~\ref{fig:nufloor-v} (vector) and  figure~\ref{fig:nufloor-s} (scalar). We show explicitly the various contributions: the recoil spectrum produced by DM events only (green), by the standard CNSN (black), by  the non-standard CNSN due either to the vector or scalar mediator (blue). In red we show the combined spectrum. In both cases, one would be able to discriminate the spectrum due to DM plus SM $\nu$ events (orange curve) from only CNSN   events (black). However, if there is an extra contribution from non-standard interactions, increasing the neutrino background (blue), one cannot discriminate anymore this situation from the total spectrum which also contain DM events(red). Both points were chosen in a region where solar neutrinos dominate the background and are only achievable for a very low energy threshold. For the nominal threshold of the LUX-ZEPLIN experiment only the vector scenario will affect the sensitivity of the experiment for $\sigma_{\chi n} \lsim$ few 10$^{-47}$ cm$^{2}$, we do not present here our results for DARWIN as they are qualitative similar to those of LUX-ZEPLIN. 

\section{Conclusions}\label{sec:conc}

Coherent neutrino scattering off nuclei is bound to become an irreducible background for the next generation of dark matter direct detection experiments, since the experimental signature is very similar to DM scattering off nuclei. In this work we have considered the case in which new physics interacts with both DM and neutrinos. In this situation, it becomes important to compute the neutrino floor while taking into account the contributions from exotic neutrino interaction. This sets the true discovery limit for direct detection experiments instead of a background-free sensitivity. For definitiveness, we have focused on two simplified models, one with a vector and one with a scalar mediator interacting with the DM and the SM particles. We calculated the bounds on the parameter space of the two simplified models imposed by the latest LUX data. These are presented in Figs.~\ref{fig:V_lim} and~\ref{fig:S_lim}.

The most interesting case is, however, the one in which some signal could be detected in a future DM direct detection experiments. In this case our models predict modifications to the standard neutrino floor. The main result of our analysis is shown in Figs.~\ref{fig:nufloor-v} and \ref{fig:nufloor-s}, in which we show that it is possible to find points in the parameter space of the models in which not only the number of events produced by DM and by the modified CNSN are compatible, but in which also the spectra are very similar. This immediately implies that the modified CNSN can mimic a DM signal above the standard neutrino floor, challenging the interpretation of a DM discovery signal. We show that the problem is more significant for experiments that can probe $m_\chi < 10$ GeV or $\sigma_{\chi n} \lsim  10^{-47}$ cm$^2$. Although a new scalar interaction will not, in practice, affect the discovery reach of future experiments such as LUX-ZEPLIN or DARWIN, a new vector interaction can mimic DM signals in a region above the standard neutrino floor of those experiments, challenging any discovery in this region.

It should be noted that the scenarios considered here lead to a variety of signatures apart from a modification of the CNSN at direct detection experiments. First and foremost, we did not account for any relic density constraints from DM annihilation. Throughout the analysis we have assumed that the DM relic density is satisfied. Secondly, the DM annihilation to neutrinos will generate signals at indirect detection experiments which will lead to additional constraints on the parameter space. Direct production of DM particles at the LHC, constrained by monojet searches will also be an additional signature of interest. Finally, exotic neutrino interactions themselves are constrained by several neutrino experiments and should be taken into account for a more complete analysis.

Despite these possible extensions of the study, our analysis is new in the sense that it considers for the first time the combined effect of exotic neutrino and DM interactions at the direct detection experiments. We demonstrate the current limits on the combined parameter space for the DM and neutrino couplings and finally demonstrate the reach of direct detection experiments.

\begin{acknowledgments}
We are thankful to Achim G\"utlein for several very useful discussions about neutrino floor calculations. We also would like to thank Geneviève Bélanger for helpful discussions. SK wishes to thank USP for hospitality during her visit, where this work originated. SK is supported by the `New Frontiers' program of the Austrian Academy of Sciences. This work was supported by Funda\c{c}\~ao de Amparo \`a Pesquisa do Estado de S\~ao Paulo (FAPESP) and Conselho Nacional de Ci\^encia e Tecnologia (CNPq).
This project has received funding from the European Union’s Horizon 2020 research and innovation programme under the Marie Sklodowska-Curie grant agreement No 674896.
\end{acknowledgments}

\bibliographystyle{JHEP}
\bibliography{neutrino_floor}
\end{document}